# Evidences for a pseudo scalar resonance at 400 GeV
# Possible interpretations

François Richard[1]

Université Paris-Saclay, CNRS/IN2P3, IJCLab[2], 91405 Orsay, France

______________________________________________________________________

**Abstract:** *This paper intends to collect the various evidences observed by ATLAS and CMS within searches for heavy scalars and pseudocalars. These searches in tt, hZ, ττ and 2jets+W, obtain individual excesses in five channels, each at a modest level of significance, ~3 standard deviations, but, put together, give a strong evidence for a pseudoscalar at ~400 GeV. Preliminary interpretations are given which suggest that additional observations should appear in the HL-LHC phase.*

## Introduction

In [1], I have reviewed the various indications for scalar and pseudoscalars shown by ATLAS and CMS. This is an ongoing task since a large fraction of the data has not yet been analysed. The purpose of this work was to examine the present situation and what can be expected from the present LHC data, from HL-LHC and from future e+e- colliders under consideration.

The present conclusion was unexpected. One signal is really sticking out, since it is observed in five channels, at a mass ~400 GeV. Although each of these observations has a significance at the 3 standard deviation level, the combination of them gives a significance far above the fatidic 5 sd level, even taking into account the 'look elsewhere' criteria. More precisely, from the following table, the product of probabilities amounts to $7 \cdot 10^{-15}$. When decreasing by 10% the number of s.d., a crude estimate for their uncertainties (numbers in parenthesis), this probability increases to $3.6 \cdot 10^{-13}$.
If one takes into account the 'look elsewhere effect', this figure increases to $5.0 \cdot 10^{-11}$, which is still below a **6 standard deviation effect**, which would correspond to a probability of $10^{-9}$.

| Reaction | Mass GeV | Significance s.d. | Probability % | Luminosity fb-1 |
|---|---|---|---|---|
| X(400)->tt | 400 | 3.5 (3.15) | 0.02 (0.08) | 36 CMS [2] |
| X(400) ->ττ | 400 | 2.2 (2) | 1.4 (2.3) | 139 ATLAS [3] |
| X(400)->ττ+b | 400 | 2.7 (2.4) | 0.35 (0.8) | 139 ATLAS 36 [3] |
| A(400)->h(125)Z | 440 | 3.6 (3.2) | 0.01 (0.07) | 36 ATLAS [4] |
| X(400) + high pt e/μ | 400 | 3 (2.7) | 0.13 (0.35) | 139 ATLAS [5] |

______________________________________________________________________

[1] richard@lal.in2p3.fr
[2] Laboratoire de Physique des 2 Infinis Irène Joliot-Curie

The hZ final state allows to identify unambiguously a **CP-odd resonance**. For what concerns the tt analysis, its authors also favour a CP-odd scalar, quote [2]: *"The largest SM background is observed for a pseudoscalar Higgs boson with a mass of 400 GeV and a total relative width $\Gamma/M$ of 4%, with a local significance of 3.5±0.3 standard deviations"*.

Hence, in the following, I will call this resonance A(400) and take it for granted.

The last channel of above list can be interpreted [1], without direct proof, as a heavy charged scalar cascading into A(400)+W, with W decaying leptonically and A(400) decaying into two jets. These two jets are effectively produced in the hZ and tt final states, which give the largest contributions, as we shall see.

Since a CP-even scalar H is observed at 660 GeV [1], with more than 4 standard deviations, in the ZZ mode into four charged leptons, one can assume that there could be a charged scalar at a similar mass and that one has H±(660)->A(400)+W, which would explain the inclusive search results from ATLAS.

As was already mentioned in [1], one should give up the usual MSSM interpretation of these scalars. The main reason comes from observing that, in this model, A(400)->hZ and H(660)->ZZ should vanish in the so called decoupling or alignment approximations, a requirement to get the SM couplings for h(125). This behaviour can, for instance, be interpreted by saying that A(400) and H(660) are iso-singlets. Something radically different is therefore happening here, irreducible to MSSM, and I will come back to what could be an explanation of this phenomenon in the next section.

In the second section, I will recall, in more detail than in [1], some of these observations and try to reach a quantitative interpretation of the A(400) properties.

Then I will try to sketch the consequences for the HL-LHC program.

## Phenomenological speculations

Before engaging in a quantitative interpretation, let me recall what can be said if we give up the standard picture of elementary scalars, the SM or SUSY paradigms.

In [1], it was assumed that **compositeness** was at work to interpret the origin of the Higgs boson discovered at LHC. There are however many avenues towards compositeness.

The most naïve one is to assume bound states of a new type of elementary particles called partons, which are the constituents of the SM particles.

Without explicitly defining the particles composing the Higgs boson, one can, on general grounds, invoke a dynamical mechanism which describes h(125) as a **pseudo-Nambu-Golstone boson**, pNGB, originating from a broken symmetry in a composite world operating at a much higher scale. This particle would therefore play the role of **pions** within QCD, the lightest bound states resulting from chiral symmetry breaking.

Alternatively, influenced by the heaviness of the top quark, which has the largest Yukawa coupling to the Higgs boson, one has speculated that the Higgs boson is itself a bound state of top quarks or from many top quarks, up to 6 in [6]. Gravitation has also been invoked to predict a very large set of scalar bound states [7]. I will not discuss any further these composite scenarios.

As for QCD with pions, there is an obvious **next step** in a composite picture. If Higgs bosons like h(125) are composite particles, it is conceivable that they can form a bound state like h-h or even h-Z and h-W, assuming that the bosons Z/W, intimately related to the BEH mechanism, are also composite. This could generate neutral scalar H (h-h) or pseudoscalar A (h-Z) resonances. This could also generate charged scalars H± by h-W bonding.

As recalled in ref [8], one even speculates that this mechanism could happen within the SM, recalling the standard case of positronium, which does not require new strong forces. Admittedly these SM resonances, not observed so far, could be too wide to be detected but one needs to be sure that this the case to avoid further misinterpretations.

Leaving the SM picture and speculating that that there are new composite force operating at the ~TeV scale, some authors provide an explicit mechanism to generate an '**Higgsinium**' bound state, [9] and [10], not necessarily very wide, in a mass region ranging between 450 to 650 GeV. Generalizing these ideas, one then gets, presumably, an interpretation of A(400) as a h-Z bound state and of H(660) as a A(400)-Z bound state.

Note that since these states are observed with a sizeable cross section at LHC, they are presumably produced by gluon-gluon annihilation, as the SM h(125), through a top loop, therefore they will presumably **decay into top pairs**. Due to the interference effects discussed in [1], these top pair final states are less easy to observe than in bosonic pairs like hh or hZ or in a fermion pair like ττ.

One can speculate that A(400)W would form a charged state at about the same mass as H(660), which would explain the cascade observed by ATLAS, X(660)±->A(400)+W, in its inclusive search. Other combinations are of course possible but have so far not been detected. In particular H(400) is not observed in the ZZ nor in the hh modes. In top pairs, CMS favours a CP-odd interpretation of the 400 GeV resonance. One should however remember that, at 400 GeV, the top pair decay of H(400) would be suppressed by a p-wave factor. Therefore, it is not possible to make a definite conclusion about its non-existence.

## Quantitative interpretation of the A(400) decays

While the ττ and hZ modes are relatively straightforward to interpret, the top pair final pair does not come out simply, due to interference effects with the gg->tt background [1].

Let us start by repeating the CMS statements [2] : "*The largest deviation from the SM background is observed for a pseudoscalar Higgs boson with a mass of 400 GeV and a total relative width Γ/M of 4%, with a local significance of 3.5±0.3 standard deviations.*"

The following table shows the results which can be deduced from the available data for different values of gAtt. Taking into account uncertainties, one can accommodate a range of Att couplings.

| gAtt | tt % | miss % | hZ % | ττ % | bb % | gg % | σA fb | Γtot GeV | Γres/M % |
|---|---|---|---|---|---|---|---|---|---|
| 1 | 42 | 46 | 3.7 | 0.55 | 0.3 | 0.04 | 9000 | 28 | 7 |
| 0.9 | 52 | 40 | 6.5 | 0.7 | 0.35 | 0.04 | 7300 | 18.3 | 5 |
| 0.8 | 66 | 19 | 5.7 | 0.87 | 0.45 | 0.06 | 5700 | 11.4 | 3.5 |
| 0.7 | 86 | 5 | 7.5 | 1.1 | 0.56 | 0.07 | 4400 | 6.7 | 2.6 |

A first question arises: Is the 4% width compatible with a CP-odd A(400), knowing that [2] quotes a ~2% two jet mass resolution width ? For mA=400 GeV, gAtt=0.9 and mt=173 GeV, the table gives $\Gamma_A$=18.3 GeV, hence $\Gamma_{res}$/M~5%. This result is at variance to what would be obtained with a CP-even Higgs which, in a p-wave, would give a width four times smaller.

This result seems to match well the conclusions of CMS but one should not forget that since:

- One is operating near threshold, the tt mode is not necessarily dominating, in fact for gAtt=1, the branching ratio in top pairs is 52%
- One should take into account the hZ channel and the $\tau\tau$ mode, produced by gg->A->$\tau\tau$ and gg->A->bb$\tau\tau$
- gAtt may differ from 1. The CMS result just sets a limit gAtt<1, with an expected limit gAtt<0.5. The difference comes from the presence of an excess. One is therefore entitled to decrease gAtt from 1 to 0.5 as I did in the table

For gAtt=1 one sees that ~50% of the modes are missing, meaning that a major decay would be missing. Which one?  The situation improves by lowering gAtt to 0.7, leaving 5% unexplained. This seems a satisfactory situation, given that we know [1] that there could be another scalar candidate at 96 GeV which would give A(400)->h(96)Z.

Taking **gAtt=0.8** as a solution which satisfies CMS observations without leaving too much missing modes, one has **gA$\tau\tau$=0.11, gAbb=0.09.**

These results do not correspond to an MSSM situation, which would say that with tan$\beta$~1.25 to 'explain' gAtt=0.8, one has gA$\tau\tau$=0.013 and gAbb=0.022, which shows an **anomalously large coupling to $\tau$** with respect to a MSSM behaviour.

## Inclusive production of A(400)

I assume that there is a decay H±(660)->A(400)+W+b, with W->e$\nu$,$\mu\nu$. One also has H(660)->A+Z with Z->ee,$\mu\mu$, which contributes 3 times less, given the Z branching ratio into leptons. Furthermore, one assumes that the charged scalar, which is produced through the processes gb->H-t and gg->H-tb, provides a spectator top which also gives an additional W for tagging.

Since ATLAS observes a signal with a cross section ~80±30 fb, which includes an acceptance factor excess of ~300±100 fb to interpret the resulting cross section. This cross section seems to be on the high side, although there is a large uncertainty due to our ignorance of the H±(660)tb coupling.

## Search for four top quarks

The two decays H±(660)->A(400)+W+b and A(400)+W+tb, with A(400)->tt should populate similar final states similar to the tttt final states searched by ATLAS [13] and CMS [14]. The latter has analysed the largest sample and reached the following results:

12.6+5.8-5.2 fb with a SM prediction of 12+2.2-2.5 fb

This result allows to set a 95% C.L. upper limit on the cross section of 25 fb ttH/A with H/A decaying into top pairs.

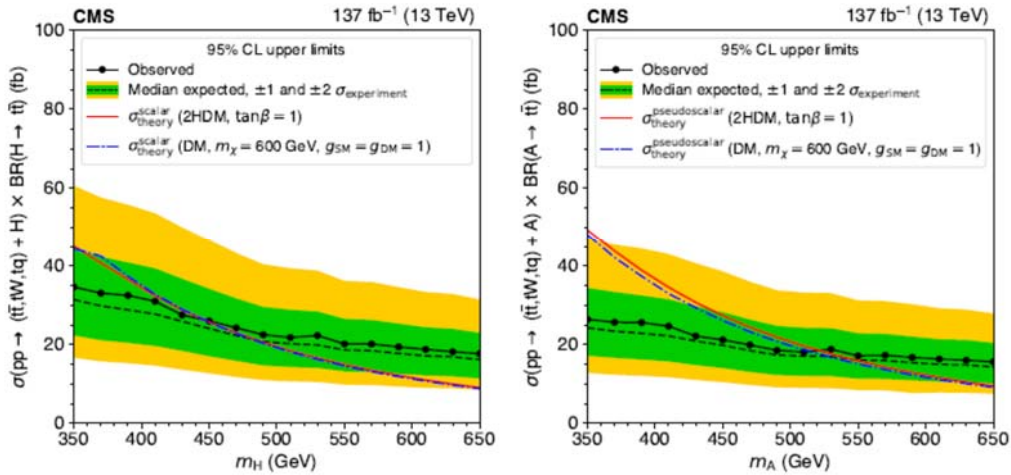

*Figure 1 : The observed (points) and expected (dashed line) 95% CL. upper limits on the cross section times branching fraction to top pairs of a new heavy scalar (left) and pseudo scalar (right), as a function of mass from the four top analysis of CMS.*

Since this analysis uses like-sign leptons and multi-b jets and therefore one needs to correct for two effects. In about 50% of the cases A(400) does not decay into top pairs which pushes the CMS limit to 50 fb. In addition the mode tt+W+b delivers twice less like sign leptons than the 4t mode, hence a 100 fb limit. There are also only 3b produces instead of 4b. Roughly speaking CMS sets a $\sigma$**<150 fb** 95% C.L. upper limit, to be compared to the rough guess of the inclusive search indication ~300±100 fb. One must admit that these two numbers are only in slight tension.

While these results are still rough guesses, one should state that the inclusive approaches do not provide a clear proof of the presence of a charged Higgs scalar.

## Future results

LHC results are still not fully analysed. In particular, the top result only relies on one-fourth the luminosity collected by CMS, also awaiting for ATLAS results. A(400) into hZ and $\tau\tau$ remain to be analysed by CMS.

For what concerns H(660), ATLAS can provide more data in ZZ which would allow to cross the 5 standard deviations border. Combining the two experiments, one may hope to confirm a weak indication in h-h from ATLAS, still marginal at the present level [11]. A confirmation from CMS for the inclusive search A(400)+W would also be very welcome.

Demonstrating that there is a transition H(660)->A(400)+Z appears very challenging. Chasing the charged Higgs in the tb mode seems to suffer interference problems [12] analogue to those in top pairs [1].

Under these circumstances, the data coming from HL-LHC will be very much needed.

Finally, one should not forget the h(96) puzzle [1]. Completing the CMS analysis with all available data is urgently needed as well as a more performing ATLAS analysis. If this particle is confirmed, it will bring several additional possibilities within this type of investigation. One should however be aware that it is not possible to distinguish h(96)->bb from Z->bb, which makes life difficult to separate h(96)-h(96) or h(96)-Z(96) from standard ZZ final states.

# Summary

One is probably living in an exciting time for what concerns searches for heavy scalars charged or neutral.

I hope that this type of spectroscopy cannot trivially be explained within the SM and opens an unexpected BSM window and a bright future for HEP.

**Apologies** for quoting very partially the rich literature on compositeness, which goes beyond my competences.